\documentclass[a4,11pt]{article}
\usepackage{frascatiphys2}
\usepackage{graphicx}
\usepackage[dvipsnames]{xcolor}
\usepackage[colorlinks, bookmarksnumbered, citecolor=Black, linkcolor=Black, pdfstartview=FitH, urlcolor=Black, linktocpage]{hyperref}
\usepackage[top = 3cm, bottom = 3cm, left = 3cm, right = 3cm]{geometry}

\newcommand{\arXiv}[1]{\href{http://arxiv.org/abs/#1}{arXiv:#1}}
\newcommand{\hepph}[1]{\href{http://arxiv.org/abs/#1}{#1}}

\begin{document}
\title{ 
SCALAR SINGLETS AT PRESENT AND FUTURE COLLIDERS
}
\author{
Dario Buttazzo        \\
{\em Physik-Institut, Universit\"at Z\"urich, CH-8057 Z\"urich, Switzerland} \\
}
\maketitle
\baselineskip=11.6pt
\begin{abstract}
A scalar singlet, coupled to the other particles only through its mixing with the Higgs boson, appears in several motivated extensions of the Standard Model.
The prospects for the discovery of a generic singlet at the various stages of the LHC, as well as at future high-energy colliders, are studied, and the reach of direct searches is compared with the precision attainable with Higgs couplings measurements. The results are then applied to the NMSSM and Twin Higgs.
\end{abstract}
\baselineskip=14pt

\section{Introduction}
Is the Higgs boson recently found by the ATLAS and CMS experiments the only scalar particle, or are there other Higgs-like states  around the Fermi scale? This question is of fundamental importance for particle physics, and motivates a detailed study of the phenomenology of additional scalars, as well as the prospects for their discovery at the LHC and future colliders\cite{our}.

The simplest example of an extended Higgs sector is realised adding just a real scalar field, singlet under all the known gauge groups, to the Standard Model (SM). Despite its great simplicity, this scenario is of considerable physical relevance, since it can easily arise in many of the most natural extensions of the SM -- e.g.\ the Next-to-Minimal Supersymmetric SM (NMSSM), Twin Higgs, some Composite Higgs models.

In general, such a singlet will mix with the Higgs boson. As a consequence, both physical scalar states are coupled to SM particles, hence they can both be produced at colliders and be observed by means of their visible decays. In the following, after briefly reviewing the main properties of a generic singlet-like scalar, I shall present the constraints on the existence of such a particle that arise from both direct searches and Higgs couplings precision measurements.
% of the couplings of the 125 GeV Higgs boson.

\section{General properties}
%Let us call
%\begin{equation}
%H = \left(\begin{array}{c}
%i\pi^+\\
%\displaystyle\frac{v + h^0 + i\pi^0}{\sqrt{2}}
%\end{array}\right),\qquad\qquad\qquad
%S = v_s + s^0,
%\end{equation}
%the Higgs and singlet fields, respectively, together with their vacuum expectation values (vev) $v = 246$ GeV and $v_s$.
Let us call $h$ and $\phi$ the two neutral, CP-even propagating degrees of freedom, with masses $m_h = 125.1$ GeV and $m_\phi$. They are related to the Higgs and singlet gauge eigenstates via a mixing angle $\gamma$.
%\begin{equation}
%h = h^0\cos\gamma + s^0\sin\gamma, \qquad\qquad\qquad \phi = -h^0\sin\gamma + s^0\cos\gamma,
%\end{equation}
%where $\gamma$ is the mixing angle.

In a weakly interacting theory,
%where operators of higher dimension can be neglected, 
the couplings of $h$ and $\phi$ are just the ones of a standard Higgs boson with the same mass, rescaled by a universal factor of $c_\gamma$ or $s_\gamma$, respectively. As a consequence, their signal strengths $\mu_{h,\phi}$
%in a given channel
are
\begin{eqnarray}
\mu_h &=& \mu_{\rm SM}(m_h)\times c_\gamma^2,\label{muh}\\
\mu_{\phi\to VV,ff} &=& \mu_{\rm SM}(m_\phi)\times s_\gamma^2\times \left(1 - {\rm BR}_{\phi\to hh}\right),\label{mus}\\
\mu_{\phi\to hh} &=& \sigma_{\rm SM}(m_\phi)\times s_\gamma^2\times {\rm BR}_{\phi\to hh},\label{mushh}
\end{eqnarray}
where $\mu_{\rm SM}(m)$ is the corresponding signal strength of a SM Higgs with mass $m$, and ${\rm BR}_{\phi\to hh}$ is the branching ratio of $\phi$ into two 125 GeV Higgs bosons.
The phenomenology of the Higgs system is therefore completely described by three parameters: $m_\phi$, $s_\gamma$, and ${\rm BR}_{\phi\to hh}$. The second state $\phi$ behaves like a heavy SM Higgs boson, with reduced couplings and an additional decay width into $hh$.

Notice that the mixing angle $\gamma$ and $m_\phi$ are not independent quantities, since the former has to vanish when the mass tends to infinity. Indeed,
\begin{equation}
\sin^2\gamma = \frac{M_{hh}^2 - m_h^2}{m_\phi^2 - m_h^2},
\end{equation}
where $M_{hh}$ is the first diagonal entry of the mass matrix of the scalar system before diagonalisation, which is proportional to the electroweak scale.

In the limit of large $m_\phi$, the Goldstone boson equivalence theorem sets the relations
\begin{equation}
{\rm BR}_{\phi\to hh} = {\rm BR}_{\phi\to ZZ} = \frac{1}{2}{\rm BR}_{\phi\to WW}.
\end{equation}
%The leading corrections to this relation for finite masses depend only on the vacuum expectation value of the singlet, $v_s$. Therefore, to a good approximation $m_\phi$, $M_{hh}$, and $v_s$ constitute a set of independent parameters that describe the phenomenology.
The exact formulae for the $hhh$ and $\phi hh$ couplings
%as functions of these parameters
are reported in reference\cite{our}.

\subsection{Higgs couplings}

\begin{table}[b]
\centering
\caption{ \it Current and expected precisions on Higgs couplings\cite{snowmass}.
}
\vskip 0.1 in
\begin{tabular}{|c|c|c|c|c|c|} \hline
$pp$ &  LHC8 & LHC14 & HL-LHC & HE-LHC & FCC-hh\\
\hline
$s_\gamma^2$ & 0.2 & 0.08--0.12 & 0.04--0.08 & ? & ?\\
$\left|\Delta g_{hhh}/g_{hhh}^{\rm SM}\right|$ & -- & -- & 0.5 & 0.2 & 0.08\\
\hline
\hline
$e^+e^-$ & ILC500 & ILC1000 & HL-ILC & CLIC & FCC-ee\\
\hline
$s_\gamma^2$ & 0.02 & 0.02 & $4\times 10^{-3}$ & 2--$3\times 10^{-3}$ & $10^{-3}$\\
$\left|\Delta g_{hhh}/g_{hhh}^{\rm SM}\right|$ & 0.83 & 0.46 & 0.1--0.2 & 0.1--0.2 & --\\
\hline
\end{tabular}
\label{table}
\end{table}

%\begin{figure}
%\includegraphics[width=0.45\textwidth]{triple1}\hfill
%\includegraphics[width=0.45\textwidth]{triple2}
%\caption{\it Isolines of $s_\gamma^2$ (coloured) and $g_{hhh}/g_{hhh}^{\rm SM}$ (dashed), for $v_s = 250$ GeV (left) and $v_s = -75$ GeV (right). The coloured region is excluded at 95\% CL.}
%\end{figure}

The measurement of the Higgs signal strengths provides a constraint on the mixing angle $\gamma$ through eq. (\ref{muh}).
At present, a global fit to 8 TeV LHC data constrain it to be $s^2_\gamma < 0.23$ at 95\% C.L.\cite{fit}. Projections for the reach of future hadron and lepton colliders\cite{snowmass} are listed in Table~\ref{table}.

%The 14 TeV LHC with 300 ${\rm fb}^{-1}$ of integrated luminosity will reach a sensitivity of about 0.15, and a further improvement of about a factor of two can be expected at the high luminosity phase. Much better sensitivities can be reached by future $e^+e^-$ colliders: ILC can probe values of $s_\gamma^2$ in the $10^{-2}$ ballpark, depending on the energy and luminosity configuration, while CLIC could even reach a few $10^{-3}$. A circular collider such as FCC-ee would have the best sensitivity, being able to measure deviations of $10^{-3}$.

Large modifications to the triple Higgs coupling
%are also proportional to $c_\gamma$, but their full dependence on the mixing angle is more involved than for the other Higgs couplings. As a consequence, large deviations from the SM value
can arise in some regions of the parameter space, even if the deviation in the signal strengths is moderate. Future collider experiments, and even the LHC, could in principle be sensitive to these modifications.
%: the high luminosity LHC and ILC are expected to be able to measure couplings of about 50\% the SM value, while higher energy leptonic and hadronic colliders (CLIC, the ILC upgrade, or FCC-hh) could go down to a about 10\%.
More details about Higgs couplings can be found in\cite{our}.

%Figure~\ref{couplings} shows isolines of the deviation in Higgs signal strengths and of the triple Higgs coupling, normalised to the SM value, in the $m_\phi$--$M_{hh}$ plane, and for two different values of $v_s$.

\begin{figure}[t]
\includegraphics[width=0.5\textwidth]{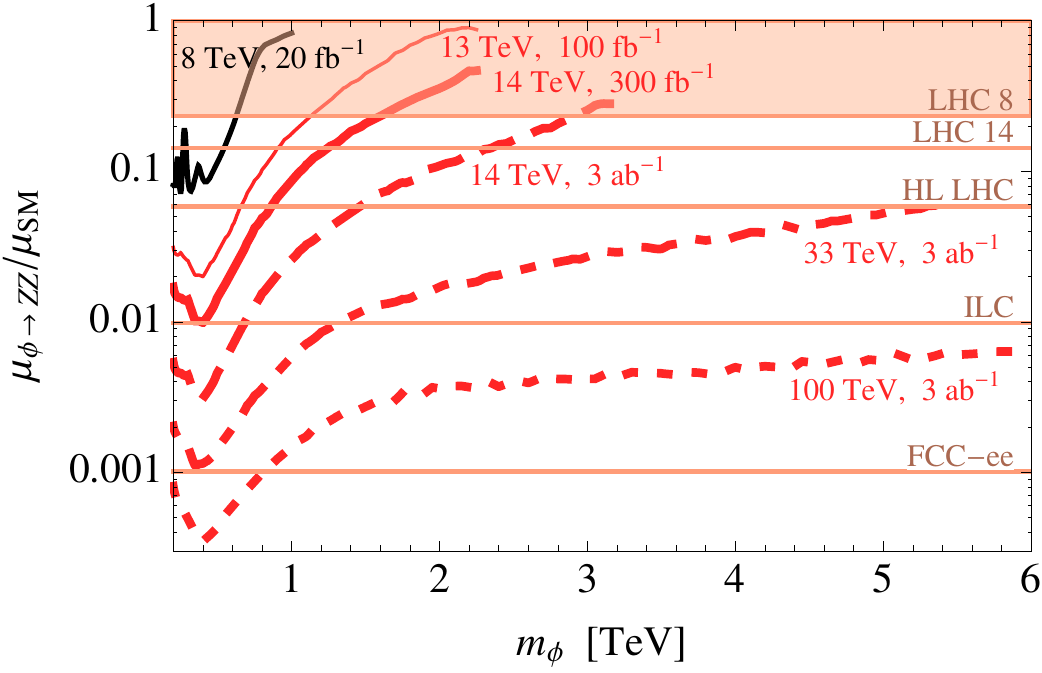}\hfill
\includegraphics[width=0.5\textwidth]{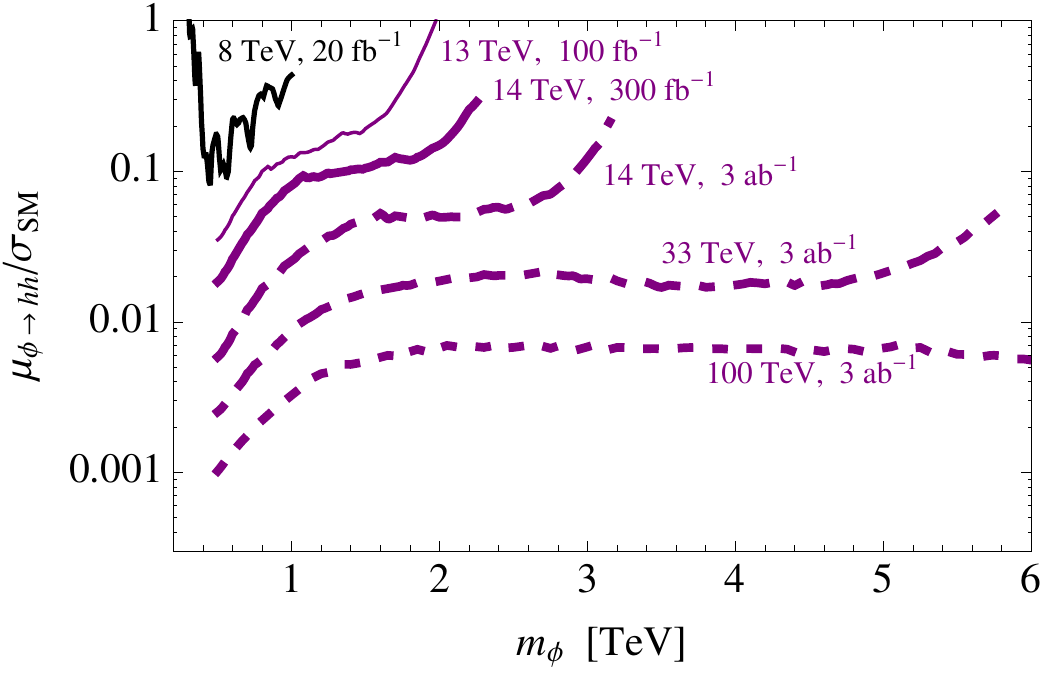}
\caption{\it Excluded values and projected reach for $\mu_{\phi\to ZZ}$ (left) and $\mu_{\phi\to hh}$ (right). In the left panel, the $s_\gamma^2$ exclusion from Higgs couplings is also superimposed, assuming a $100\%$ branching ratio into vectors.\label{direct}}
\end{figure}

\section{Direct searches}

\begin{figure}[t]
\includegraphics[width=0.45\textwidth]{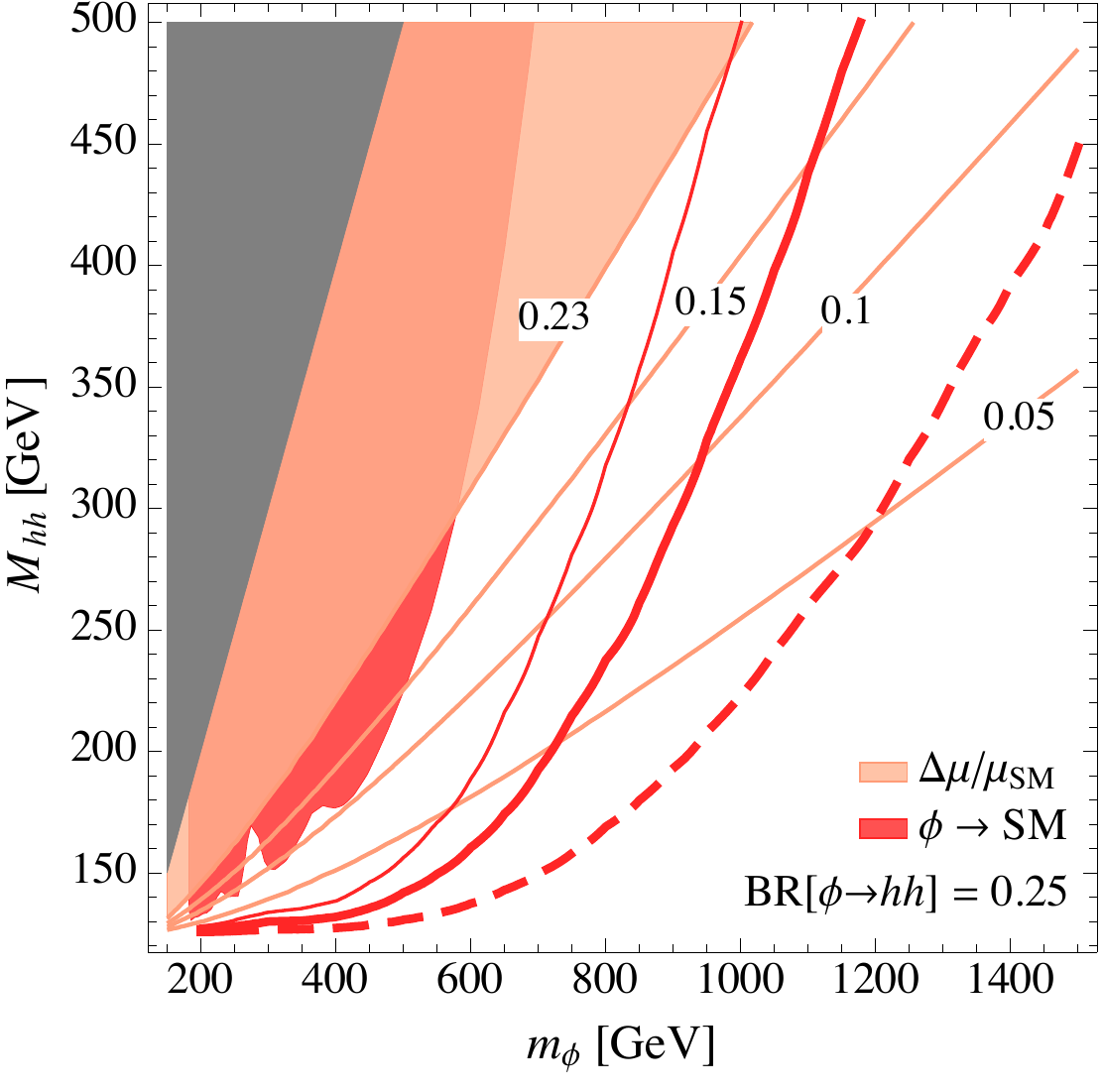}\hfill
\includegraphics[width=0.464\textwidth]{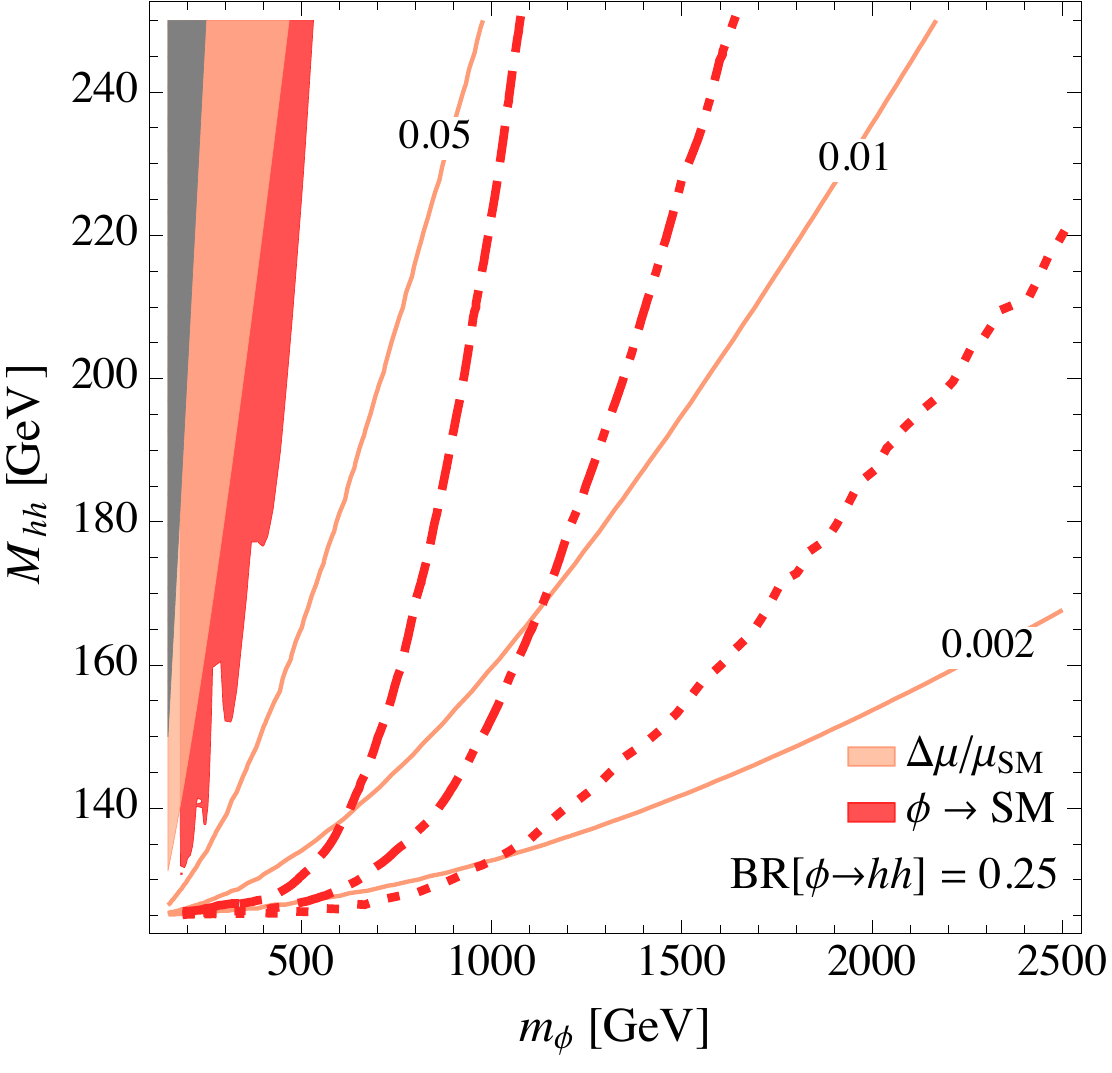}
\caption{\it Comparison between the combined reach of direct searches and Higgs coupling measurements, in the plane $m_\phi$--$M_{hh}$. ${\rm BR}_{\phi\to hh}$ has been fixed to 0.25 for simplicity. Left: region relevant for the LHC. Right: projections for future colliders. The notation for the lines is the same as in Figure~\ref{direct}.\label{all}}
\end{figure}

The main decay channels of a heavy singlet are into a pair of $W$ and $Z$ vector bosons, or into a pair of Higgs bosons, if kinematically allowed.

Both the ATLAS and CMS collaborations provide a combined limit from all the $WW$ and $ZZ$ channels\cite{ZZ}, with the strongest bound always coming from searches in the $4\ell$ and $2\ell 2\nu$ final states. In the di-Higgs channel, the main constraint comes from the $4b$ final state\cite{hh}.
%, but $2b2\gamma$ is also important for low masses.
All these searches are already sensitive to cross-sections smaller than the ones for a SM Higgs at the same mass, and exceed the reach of Higgs coupling measurements for low enough~$m_\phi$.
% resonance masses.

Projections for future colliders have been obtained in\cite{our}, rescaling the expected limits from the 8 TeV LHC with the parton luminosities of the backgrounds, following the procedure presented in\cite{ttw}.
%This method is subject to a number of rather strong assumptions and simplifications, but is nevertheless suited for obtaining a quick estimate of the reach in cross-section with a reasonable level of accuracy.
The colliders that have been considered are: the 8 TeV, 13 TeV, and 14 TeV LHC, its high-luminosity upgrade, a possible 33 TeV energy upgrade, and a futuristic 100 TeV FCC-hh.

Figure~\ref{direct} shows the present and extrapolated limits on the $\mu_{\phi\to VV}$ and $\mu_{\phi\to hh}$ signal strengths, normalised to SM values of the cross-sections.
%The sensitivity in the two channels is similar.
In the left panel the projections for 125 GeV Higgs couplings measurements are also shown, in the limit of small ${\rm BR}_{\phi\to hh}$. Figure~\ref{all} again shows a comparison between direct and indirect searches, but this time in the $m_\phi$--$M_{hh}$ plane, and for ${\rm BR}_{\phi\to hh} = 1/4$. The direct exclusion is dominated by $\phi\to VV$.

%In the limit of small ${\rm BR}_{\phi\to hh}$, $\mu_{\phi\to VV}\simeq s_{\gamma}^2 \mu_{\rm SM}$, so in the left-handed plot the projections for the 125 GeV Higgs couplings measurements are also shown.
%One can see that for masses below about a TeV, direct searches always give a stronger bound, for comparable experimental setups.
%Figure~\ref{all} again shows a comparison between direct and indirect searches, but this time in the $m_\phi$--$M_{hh}$ plane.
%The direct exclusion is dominated by $\phi\to VV$ in virtue of the smaller branching ratio into $hh$, which has been fixed to its asymptotical value of $1/4$ everywhere.

%sdfsdfsd

\section{Explicit models}
\subsection{Supersymmetry}

The Higgs sector of the NMSSM\cite{ellwanger} contains the two usual doublets $H_{u,d}$, plus a singlet scalar $S$, coupled through a Yukawa interaction $\lambda H_u H_d S$ in the superpotential. An extra contribution to the Higgs mass is generated at tree-level by $\lambda$, and reduces the size of the radiative correction needed to obtain $125$ GeV. At the same time, the fine-tuning of the electroweak scale $v$ is reduced.
%The two main motivations for such a scenario are:
%\begin{itemize}
%\item an extra contribution to the Higgs mass at tree-level, which reduces the size of the radiative correction needed to generate $m_h = 125\, {\rm GeV}$,
%\item an overall reduction of the tuning of the electroweak scale $v$.
%\end{itemize}

In the decoupling limit for the heavy doublet, the CP-even states are the SM Higgs and the singlet, and can be matched to the previous scenario via\cite{nmssm}
\begin{equation}
M_{hh}^2 = m_Z^2 c_{2\beta}^2 + v^2\lambda^2 s_{2\beta}^2 + \Delta^2,
\end{equation}
where $\Delta$ is the radiative correction and $\tan\beta = v_u/v_d$.
Figure~\ref{models} (left) shows the current exclusions and projections from both direct searches and Higgs couplings, in the plane $m_\phi$--$\tan\beta$, for fixed values of $\lambda = 1$ and $\Delta = 70$ GeV.

\subsection{Twin Higgs}

\begin{figure}
\includegraphics[width=0.45\textwidth]{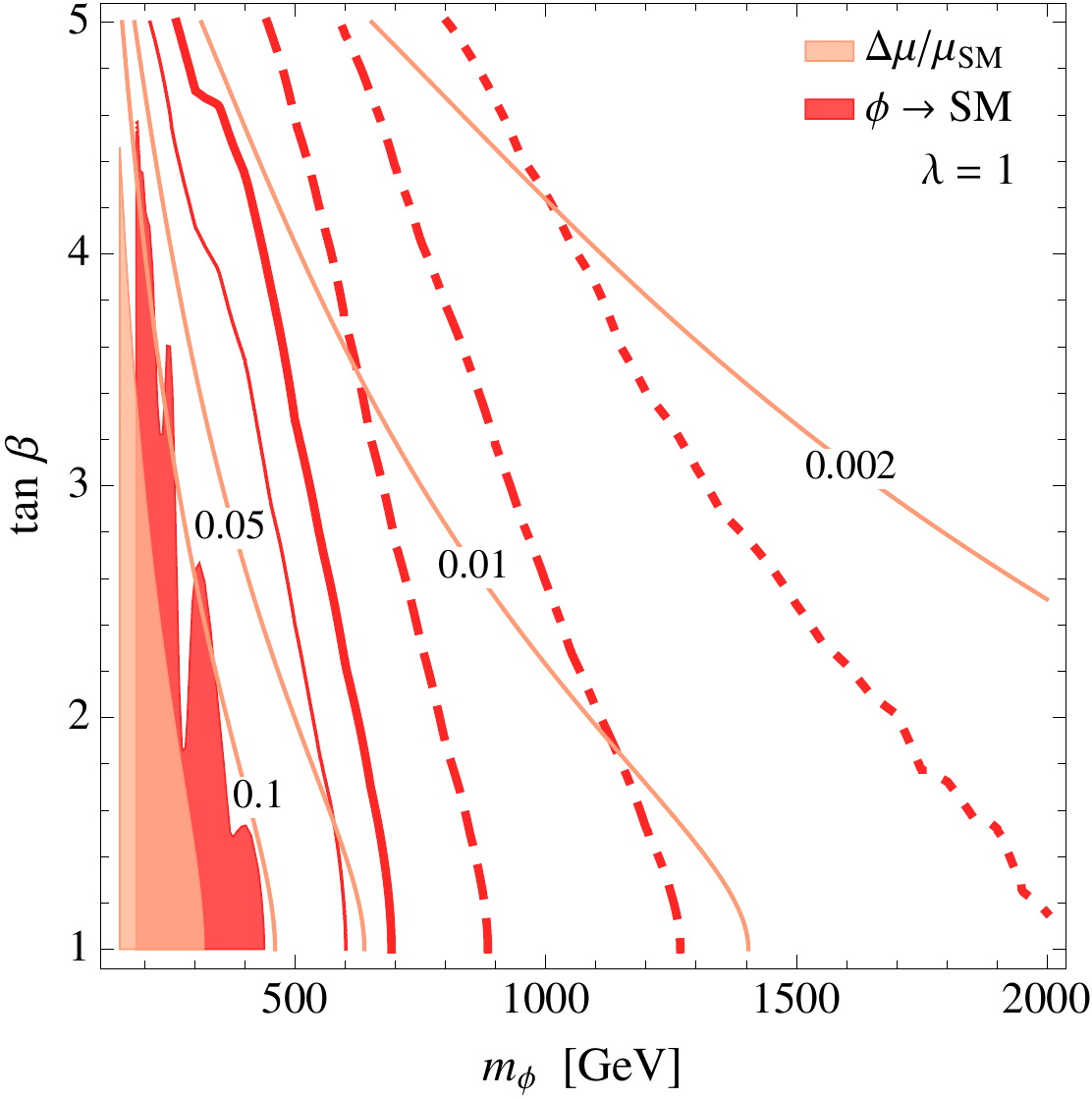}\hfill
\includegraphics[width=0.482\textwidth]{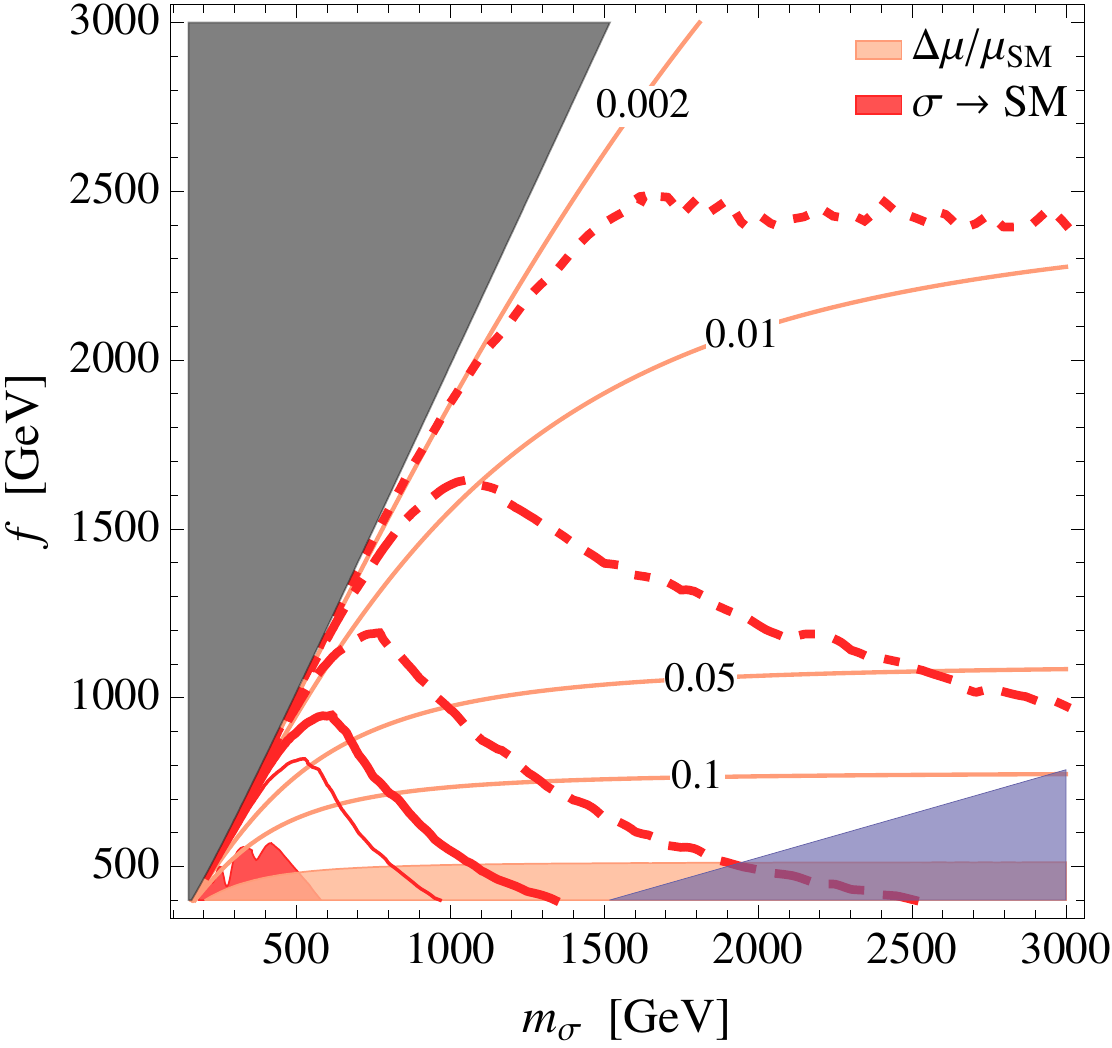}
\caption{\it Current exclusions and projections for the NMSSM singlet with $\lambda = 1$ and $\Delta = 70$ GeV (left), and the Twin Higgs radial mode (right). The notation is the same as in Figure~\ref{direct}. In the purple region the width $\Gamma_\phi > m_\phi$.\label{models}}
\end{figure}

In Twin Higgs models\cite{th},
%, the Higgs is a pseudo-Goldstone boson of a spontaneously broken symmetry. Compared to other scenarios of this type,
a naturally light Higgs is obtained without the presence of coloured particles close to the TeV scale. This is achieved introducing a copy of the SM field content and gauge symmetries, ${\rm SM_A}\times {\rm SM_B}$. The Higgs potential has an approximate global ${\rm SO}(8)$ symmetry, which is spontaneously broken at a scale $f$, and the Higgs $h = H_{\rm A} \cos\gamma + H_{\rm B}\sin\gamma$ is a Goldstone boson of this breaking. Quadratic ``divergences'' in the Higgs mass cancel between the A and B sectors, while all the new Twin particles
%-- including the $H_{\rm B}$ scalar --
are SM singlets.

The phenomenology of the ``radial mode'' $\sigma = H_{\rm B}\cos\gamma - H_{\rm A}\sin\gamma$
%is an admixture of SM doublet and singlet, and its phenomenology
is described by eq.~(\ref{mus}), (\ref{mushh}). The mixing angle is proportional to $v/f$, and one has
\begin{equation}
M_{hh}^2 = \frac{v^2}{f^2}(m_\sigma^2 + m_h^2).
\end{equation}
The only difference with respect to the previous cases is the presence of an invisible width into $W_{\rm B}$ and $Z_{\rm B}$ bosons.
Figure~\ref{models} (right) illustrates the present and future constraints in the plane $m_\sigma$--$f$, which are the only two free parameters of the model. One can see that direct searches for the radial mode are the most powerful probe for a Twin Higgs scenario, at least for not too large values of $m_\sigma$ and $f$.

\section{Conclusions}
Searches for scalar singlets at colliders can be an important probe for the extended Higgs sectors of many physically motivated models, and complementary to the measurement of Higgs couplings. By means of only three parameters that determine the phenomenology in a completely general way, the reach of future colliders in the relevant $VV$ and $hh$ channels has been studied. On the other hand, already the second run of the LHC can efficiently explore this scenario, and will provide valuable information in the near future.

\section*{Acknowledgements}
I thank Filippo Sala and Andrea Tesi for the stimulating collaboration on which the results presented here are based. This research was supported in part by the Swiss National Science Foundation (SNF) under contract 200021-159720.

\end{document}